# Energy Transfer Mechanisms and Equipartitioning in non-Equilibrium Space-Charge-Dominated Beams

R. A. Kishek , P. G. O'Shea, and M. Reiser,
Institute for Plasma Research, U. Maryland, College Park, MD 20742, USA

*Abstract*

A process of energy transfer is demonstrated in non-equilibrium charged particle beams with anisotropy and space charge. Equipartitioning of energy between available degrees of freedom occurs in just a few betatron wavelengths, without halo formation. Collective space charge modes similar to those observed in recent experiments provide the underlying coupling mechanism. Since laboratory beams are commonly far from equilibrium, the traditional K-V stability analysis does not necessarily apply, implying that selection of an operating point based on theory does not necessarily avoid equipartitioning. Furthermore, the rate of equipartitioning is shown to depend on a single free parameter related space charge content of the final (equipartitioned) beam, and does not depend on how the kinetic energy is initially distributed between the two planes.

Modern designs for high intensity linear accelerators are frequently based on the presence of large anisotropies between the longitudinal and transverse directions [1-2]. Since the time scale for Coulomb collisions (intrabeam scattering) is long relative to the size of the machine, thermodynamic equipartitioning based on particle collisions is usually ignored. Of more significance to the designer is the collective space charge interactions which may under some circumstances couple the degrees of freedom and allow the energy transfer. The question which has preoccupied beam physicists for the past two decades [1-6] is precisely under what conditions does a collisionless beam equipartition?

The wealth of theoretical studies in the past 2 decades has contributed much to our understanding, but the overall picture remains far from complete. Thermodynamic considerations [1, 6] have been used to predict the final equilibrium state, but are unable to address the detailed energy transfer mechanism or the timescales involved. As evidenced by some computer simulations, unstable space charge modes have been advanced as a likely mechanism [3-4]. Using the same framework that Gluckstern has used for an isotropic beam [7], Hofmann has analytically derived the stability properties of anisotropic distributions to small perturbations [8]. The derivation assumes a Kapchinskij-Vladimirskij (KV) -like but anisotropic equilibrium distribution in a uniform focusing channel, so as to make the mathematics manageable, and results in charts delineating stable and unstable regions that can be useful to the accelerator designer, if correct. The theory has been tested with simulations having KV and waterbag initial distributions [3]. The problem is that beams in real machines are usually quite far from equilibrium, and certainly not a KV. We therefore need to ask to what extent are the features of such stability charts a result of the choice of KV distribution?

To address this issue, we self-consistently simulate anisotropic beams using the particle-in-cell code WARP [9] and using non-equilibrium initial distributions to model realistic beams. The lack of equilibrium in the initial distributions makes comparison to KV stability theory more difficult. It also implies, as we shall see, that certain space charge modes are born from the initial mismatch of the distribution. In addition to examining the energy transfer mechanism, we explore its scaling and demonstrate that the rate of energy transfer depends on a single free parameter related to the ratio of space charge forces to external focusing forces.

In many respects, the simulations presented here are very similar to those of a recent experiment, albeit in an isotropic system, by Bernal, et. al. [10]. The space-charge-dominated electron beam in that experiment exhibited wave-like density modulations which were traced to the lack of detailed equilibrium at the source. Simulations with the WARP starting with a semi-Gaussian (SG) distribution [uniform density and Gaussian velocity distribution with a uniform temperature across x or y] have accurately reproduced the density modulations [10]. A KV initial distribution, on the other hand, did not reproduce the experiment. In this paper we introduce anisotropy into the such simulations.

We use the 2-½ D slice version of WARP [9], which advances particles in a transverse slice under the action of external forces and the self-consistent self-fields. To simplify the issue, the external focusing is chosen to be uniform along z (and equal in x and y), resembling the focusing obtained from a uniform distribution of background ions [11]. Typically we use a $256 \times 256$ grid for the Poisson solver, a step size of 4 mm along z, and 20,000 particles, with test simulations up to 400,000 particles. Extensive testing of the numerics have

---

ramiak@ebte.umd.edu

demonstrated that the simulations are very robust with respect to the choice of numerical parameters.

In the simulation shown in Fig. 1, a 10 kV, 50 mA electron beam having a 7.5 mm radius is launched inside a 1" radius circular pipe. The anisotropy is introduced by fixing the external focusing strength at $k_o = 3.972$ m$^{-1}$ in both directions, and picking initially different emittances in x and y ($\varepsilon_x = 100$ µm, while $\varepsilon_y = 50$ µm, unnormalized effective). This implies that the tune depressions are different, namely $(k/k_o)_x = \lambda_{\beta o}/\lambda_{\beta x} = 0.41$; $(k/k_o)_y = \lambda_{\beta o}/\lambda_{\beta y} = 0.26$, $\lambda_\beta$ being the betatron wavelength and the subscript 'o' denoting zero space charge. The initial beam sizes in x and y are chosen to be matched solutions of the rms envelope equations, assuming the emittances will not change. Naturally, any change in emittance will also induce an rms mismatch. In subsequent simulations, we vary the ratio $\varepsilon_x / \varepsilon_y$ and also the beam current to explore the scaling of the energy transfer.

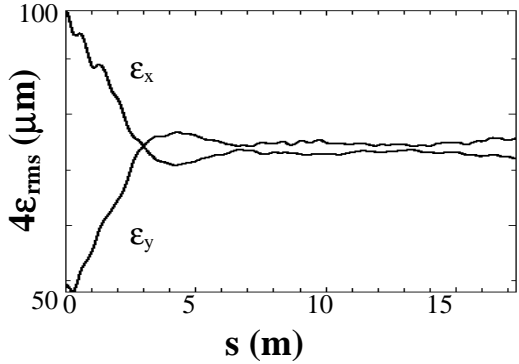

**Fig. 1:** Emittance exchange due to equipartitioning of a SG beam in a symmetric uniform focusing channel ($\alpha_o = k_{yo}/k_{xo}=1$) and starting with $\varepsilon_x = 2\varepsilon_y$; $\xi = (k/k_o)_{final} = 0.353$, T = 3.17.

Hofmann [8] has found that three dimensionless variables are needed to describe the parameter space of anisotropic beams. In this paper we choose the following combination: (i) the ratio of zero-current betatron tunes: $\alpha_o \equiv k_{yo}/k_{xo}$, which is set by the external lattice and is not a free parameter for practical purposes; (ii) the square root of the ratio of the total transverse kinetic energy to the external field energy:

$$\chi \equiv \sqrt{\frac{(e_x/a)^2 + (e_y/b)^2}{k_{xo}^2 a^2 + k_{yo}^2 b^2}},$$

also invariant, due to conservation of energy; and (iii) the initial ratio of kinetic energies in the two transverse directions: $T \equiv T_x/T_y = \varepsilon_x k_x / \varepsilon_y k_y$. We set $\alpha_o = 1$ for this paper. The 2nd parameter can be shown to approximately equal the tune depression of the final equipartitioned beam [12]. Note that the 3rd parameter is the only variable measuring the degree of anisotropy and the only one that changes as the beam equipartitions.

For the simulation in Fig. 1, $\xi = (k/k_o)_{final} = 0.35$ and T = 3.17, placing it in the space-charge-dominated regime with a fairly strong energy anisotropy. This simulation is typical of simulations started with a SG distribution in that the beam is observed to transfer energy between the two directions until the beam sizes and emittances in the two transverse planes are equal [Fig. 1]. The final emittance satisfies conservation of energy [1, 13]. Close examination of the density profile of the beam [14] reveals that the energy transfer mechanism is precisely the density oscillations that appear in the symmetric case [10]. Noting that the speed of propagation of the density crests from the edge to the center depends on the tune depression [15], the wave velocities in an anisotropic beam will be different because the tune depressions are initially different in x and y. What begins as a ring at the beam edge transforms into an ellipse with a different eccentricity, i.e., an initial temperature anisotropy translates into a density anisotropy downstream. As the wave breaks at the center of the beam, it generates higher-order modes that couple the two transverse directions and facilitate the transfer of kinetic energy which leads to equipartitioning.

Simulations started with a KV distribution evolve differently, and the KV beam does not always equipartition [14]. The KV beam remains stable unless the operating point allowed one of the instabilities predicted by the KV stability theory [8] to be excited. Simulations started with a SG beam are not as easy to reconcile with the theory, and the differences are discussed more fully in [14].

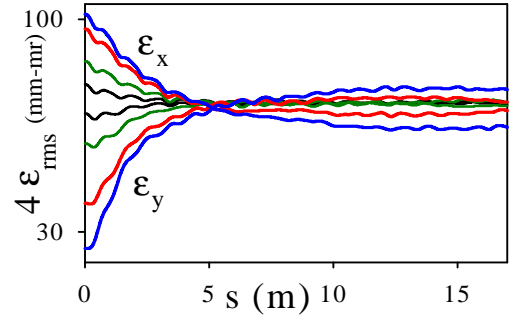

**Fig. 2:** Evolution of emittance for beams with different degrees of anisotropy; $\xi = 0.188$ and $\alpha_o = 1$ for all beams, T varies from 1 to 16.

In Fig. 2, we vary the parameter T and find that the characteristic distance (or time) over which a beam equipartitions is over a large range independent of the degree of anisotropy, provided $\alpha_o$ and $\xi$ are the same. The 2nd parameter is therefore the only one which governs the rate of equipartitioning. In other words, the rate of equipartitioning is unaffected by the way the kinetic energy is distributed in the two directions and hence depends only on the final isotropic state, as long as the total kinetic energy is held the same.

To quantify the rate of this process, we define a "characteristic distance", $s_{eq}$, over which the emittances approach their final value for the first time. In Fig. 3 the beam current is systematically varied to explore the dependence on the parameter, $\xi$. The only caveat is that the ratio of emittances is held constant, so T changes slowly because of changes in the matched beam size. Nevertheless, as just demonstrated, changes in T do not affect the equipartitioning time.

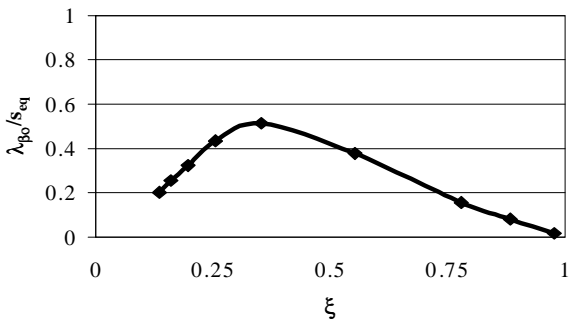

**Fig. 3:** The rate of equipartitioning, defined (see text) as $\lambda_{\beta o} / s_{eq}$, as a function of $\xi$; $T_x/T_y$ varies slowly between 2 and 4, $\alpha_o = 1$.

Three conflicting factors affect the equipartitioning rate: (a) the space charge content $(1 - \xi^2)$ since that determines the strength of the coupling; (b) the initial amplitude, and (c) the propagation speed of the perturbation, since it is the main vehicle that effects the coupling. We have a tradeoff between the amplitude of the initial perturbation (b) and the space charge content (a) since an equilibrium thermal distribution converges to a semi-Gaussian in the space charge limit where the temperature (emittance) is zero. Further, the propagation speed (c) is found in the isotropic case to peak at intermediate tune depressions [15]. The strongest coupling therefore takes place at the intermediate $\xi$ of ~ 0.3. Note that the equipartitioning distance can be as small as two zero-current betatron periods, and increases to larger values at either extreme. Even at the weak tune depression of 0.87, the emittance can change significantly in about 12 betatron periods.

In conclusion we pose a few questions on the implications of this work. While theoretical studies have been limited to small perturbations from equilibrium, it is obvious that the possibility of equipartitioning because of a large perturbation needs to be investigated, since ultimately beams in real machines will experience such perturbations. Machine designs employing unequipartitioned beams need to be carefully reexamined if space charge plays a role. Our finding that the rate of equipartitioning depends only on the tune depression of the final isotropic beam leads us to propose that a general anisotropic beam can be modeled by using an "equivalent isotropic beam" having the same ratio of total kinetic to total external energies. Work along these lines needs to be continued to explore cases where the external focusing is not symmetric ($\alpha_o \neq 1$), and to explore transverse-longitudinal equipartitioning in bunched beams.

We are grateful to A. Friedman, D. Grote and S. M. Lund for the WARP code; to S. Bernal, C. L. Bohn, I. Haber, I. Hofmann, and M. Venturini for valuable discussions; and to M. Holland for assistance with simulations. This work is supported by the U.S. Department of Energy grant numbers DE-FG02-94ER40855 and DE-FG02-92ER54178. The WARP code runs on DOE supercomputers provided by NERSC at LBNL.